\documentclass[runningheads]{svmult}

\usepackage{makeidx}   
\usepackage{graphicx}  
\usepackage{subeqnar}  
\usepackage{multicol}  
\usepackage{physprbb}  
\makeindex             

\newcommand{\la}{\mathrel{\vcenter
     {\offinterlineskip \hbox{$<$}\hbox{$\sim$}}}}

\begin{document}
\title*{Core Collapse and Then?\protect\newline 
The Route to Massive Star Explosions}
\toctitle{Core Collapse and Then? The Route to Massive 
Star Explosions}
%
%
\titlerunning{Core Collapse and Then?}
%
\author{Hans-Thomas Janka\inst{1}
\and Robert Buras\inst{1}
\and Konstantinos Kifonidis\inst{1}
\and\\ Tomek Plewa\inst{2,3}
\and Markus Rampp\inst{1}}
\authorrunning{H.-Th. Janka at al.}
%
%
\institute{Max-Planck-Institut f\"ur Astrophysik,
           Karl-Schwarzschild-Str.~1, D-85741 Garching, Germany
\and Dept. of Astronomy and Astrophysics and 
           Center for Astrophysical Thermonuclear Flashes,
           The University of Chicago, Chicago, IL 60637, U.S.A.
\and Nicolaus Copernicus Astronomical Center, Bartycka 18, 
           00716 Warsaw, Poland}

\maketitle              
 
\begin{abstract}
The rapidly growing base of observational data for supernova
explosions of massive stars demands theoretical explanations.
Central of these is a self-consistent model for the physical
mechanism that provides the energy to start and drive the
disruption of the star. We give arguments why the delayed
neutrino-heating mechanism should still be regarded as the 
standard paradigm
to explain most explosions of massive stars and show how large-scale
and even global asymmetries can result as a natural consequence of
convective overturn in the neutrino-heating region behind the 
supernova shock. Since the explosion is a threshold phenomenon and
depends sensitively on the efficiency of the energy transfer by 
neutrinos, even relatively minor differences in numerical 
simulations can matter on the secular 
timescale of the delayed mechanism. To enhance this point, we
present some results of recent one- and two-dimensional computations,
which we have performed with a Boltzmann solver for the neutrino
transport and a state-of-the-art description of neutrino-matter
interactions. Although our most complete models fail to explode, 
the simulations demonstrate that one is encouragingly close to
the critical threshold because a modest variation of the neutrino
transport in combination with postshock convection leads to 
a weak neutrino-driven explosion with properties that fulfill
important requirements from observations. 
\end{abstract}

\section{Introduction}
\label{sec:problem}

The primary energy source for powering supernovae of massive
stars is the gravitational binding energy of the newly formed
proto-neutron star or proto-black hole
(energy from nuclear reactions contributes at a minor
level). To initiate and drive the explosion, energy from some
temporary storage, e.g., internal or rotational energy of the
compact remnant, must be transferred to the outer stellar 
layers to be finally converted to kinetic
energy of the ejecta. This might be achieved by hydrodynamical
shocks, by neutrinos, or by magnetic fields as mediators.
Accordingly, one distinguishes between
\begin{itemize}
\item[(i)] the (prompt) mechanism, which works 
on a dynamical timescale by the hydrodynamical shock that
is created at the moment of core bounce,
\item[(ii)] the delayed, neutrino-driven mechanism which
starts the explosion on the secular timescale of 
neutrino-energy deposition behind the supernova shock,
\item[(iii)] and the magnetohydrodynamical (MHD) mechanism,
which requires that initial seed magnetic fields are amplified 
to a dynamically relevant strength by differential rotation.
\end{itemize}

Intense radiation or relativistic outflows of charged particles
might also play a role for very special conditions.
They may, for example, originate from the vicinity of 
an accreting black hole that has formed after the collapse
of a rotating stellar core. Relativistic jets as driving 
mechanism are currently discussed for stellar explosions that
have been observed in association with gamma-ray bursts
(see the contributions by S.~Woosley and A.~MacFadyen at
this conference).

Depending on the mediator, the
conditions for efficient energy transfer, the corresponding
timescale, and the tapped energy reservoir are different.
A lot of work has been spent in the past 40 years on the
search for a viable supernova mechanism and the study of
the various theoretical suggestions. A brief review of 
these efforts and the current status of our 
knowledge can be found in Ref.~\cite{janbur02b}.

The relevance of the different mechanisms for stellar 
explosions listed above
depends on the (poorly known) physical conditions
in collapsed stellar cores and on the properties of
the progenitor stars. Some of the involved requirements are
more likely to be fulfilled than others, some combinations of
necessary conditions may be more frequent and more typical,
while others may be realized only in rare cases and
for very special, exceptional circumstances.

The neutrino-driven mechanism~\cite{wil85,betwil85} involves 
a minimum of controversial assumptions and uncertain
degrees of freedom in the physics of collapsing stars.
It relies on the importance of neutrinos and their
energetic dominance in the supernova core. After the detection
of neutrinos in connection with Supernova~1987A and the
overall confirmation of theoretical expectations
for the neutrino emission, this can no longer be
considered as a speculative assumption but is an
established fact. Of course, this does not mean that
such a minimal input is sufficient to understand the cause
of supernova explosions and to explain all observable
properties of supernovae. But at least it can be taken as a
good reason to investigate how far one can advance with a
minimum of imponderabilities.

\section{Observational Facts}
\label{sec:observations}

Progress in our understanding of the processes that lead
to the explosion of massive stars is mainly based on elaborate
numerical modeling, supplemented by theoretical analysis
and constrained by a growing data base of observed
properties of supernovae. The latter may carry imprints from
the physical conditions very close to the center of the 
explosion. Observable features at very large radii, however, 
can be linked to the actual energy source of the explosion
only indirectly through a variety of intermediate
steps and processes. Any interpretation with respect to
the mechansim that initiates the explosion
therefore requires caution.

A viable model for the explosion mechanism of massive 
stars should ultimately be able to explain the observed 
explosion energies, nucleosynthetic yields (in particular 
of radioactive isotopes like $^{56}$Ni, which are created
near the mass cut), and the masses of the compact remnants
(neutron stars or black holes) and their connection with
the progenitor mass. 

Recent evaluations of photometric and
spectroscopic data for samples of well-observed Type-II 
plateau supernovae reveal a wide continuum of kinetic energies
and ejected nickel masses. Faint, low-energy cases seem to 
be nickel-poor whereas bright, high-energy explosions tend
to be nickel-rich and associated with more massive 
progenitors~\cite{ham02}. This direct correlation between 
stellar and explosion properties, however, is not 
apparent in an independent analysis by Nadyozhin~\cite{nad02}
who speculates that more than one 
stellar parameter (rotation or magnetic fields besides the
progenitor and core mass) might determine the explosion
mechanism. A large range of nickel masses and explosion
energies was also found for Type Ib/c supernovae~\cite{ham02}.
Interpreting results obtained by the fitting of optical 
lightcurves and spectra, Nomoto et al.~\cite{nom02} came up 
with the proposal that explosions of stars with main
sequence masses above 20--25$\,$M$_{\odot}$ split up to a
branch of extraordinarily bright and energetic events
(``hypernovae'') at the one extreme and a branch of faint, 
low-energy or even ``failed'' supernovae at the other. 
Stars with such large masses might collapse to black holes 
rather than neutron stars. The power of the explosion could
depend on the amount of angular momentum in the collapsing
core, which in turn can be sensitive to a number of effects such
as stellar winds and mass loss, metallicity, magnetic fields,
binarity or spiraling-in of a companion star in a binary system.

Anisotropic processes and large-scale mixing between
the deep interior and the hydrogen layer had to be invoked in
case of Supernova~1987A to explain the shape of the lightcurve,
the unexpectedly early appearance of X-ray and $\gamma$-ray
emission, and Doppler features of spectral lines (for a review,
see~\cite{nom94}). More than ten years after the explosion, the 
expanding debris exhibits an axially symmetric 
deformation~\cite{wanwhe02}.
Supernova~1987A therefore seems to possess
an intrinsic, global asymmetry. The same conclusion was drawn
for other core-collapse supernovae (Type-II as well as Ib/c)
based on the fact that their light is
linearly polarized at a level around 1\% with a tendency
to increase at later phases when greater depths are
observed~\cite{wanhow01,leofil01}.
This has been interpreted as evidence that the inner portions 
of the explosion, and hence the mechanism itself, are strongly
non-spherical~\cite{hoewhe99,wheyi00},
possibly associated with a ``jet-induced'' 
explosion~\cite{wanwhe02,khohoe99}. This is a very 
interesting and potentially relevant conjecture. 
It does, however, not necessarily
constrain the nature of the physical process that mediates the 
energy transfer from the collapsed core of the star to the
ejecta and thereby creates the asphericity.

%
%
%
%

\begin{figure}[htb!]
\begin{center}
\includegraphics[width=1.0\textwidth]{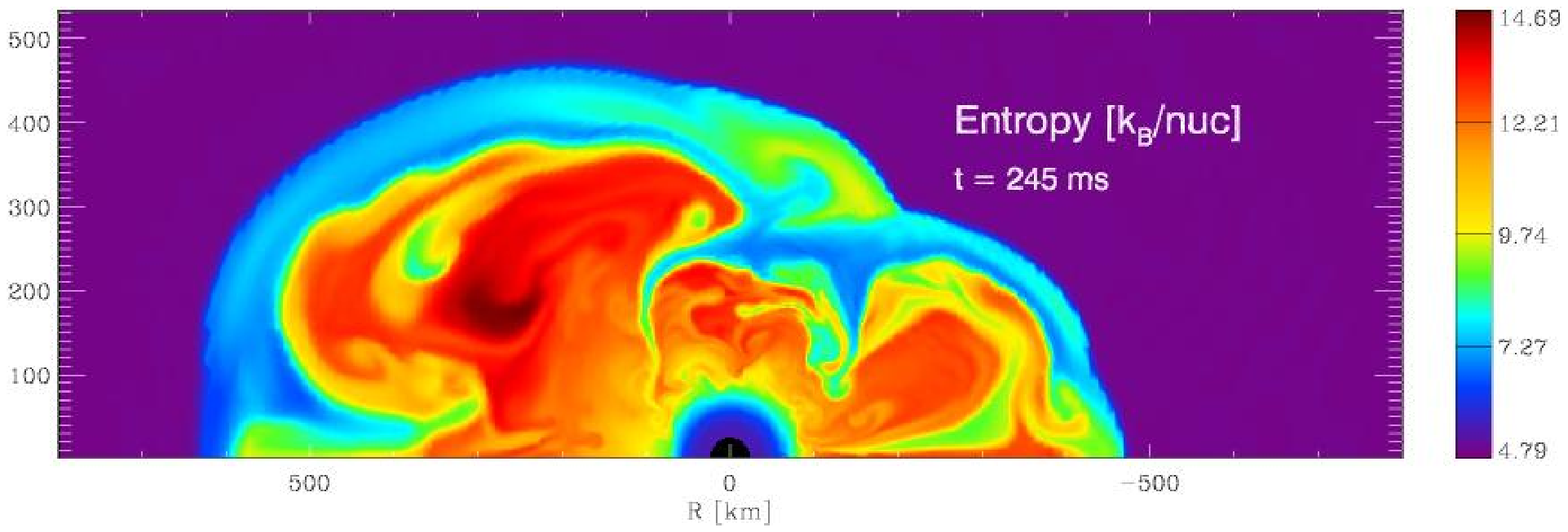}
\end{center}
\vspace{-5mm}
\begin{center}
\includegraphics[width=1.0\textwidth]{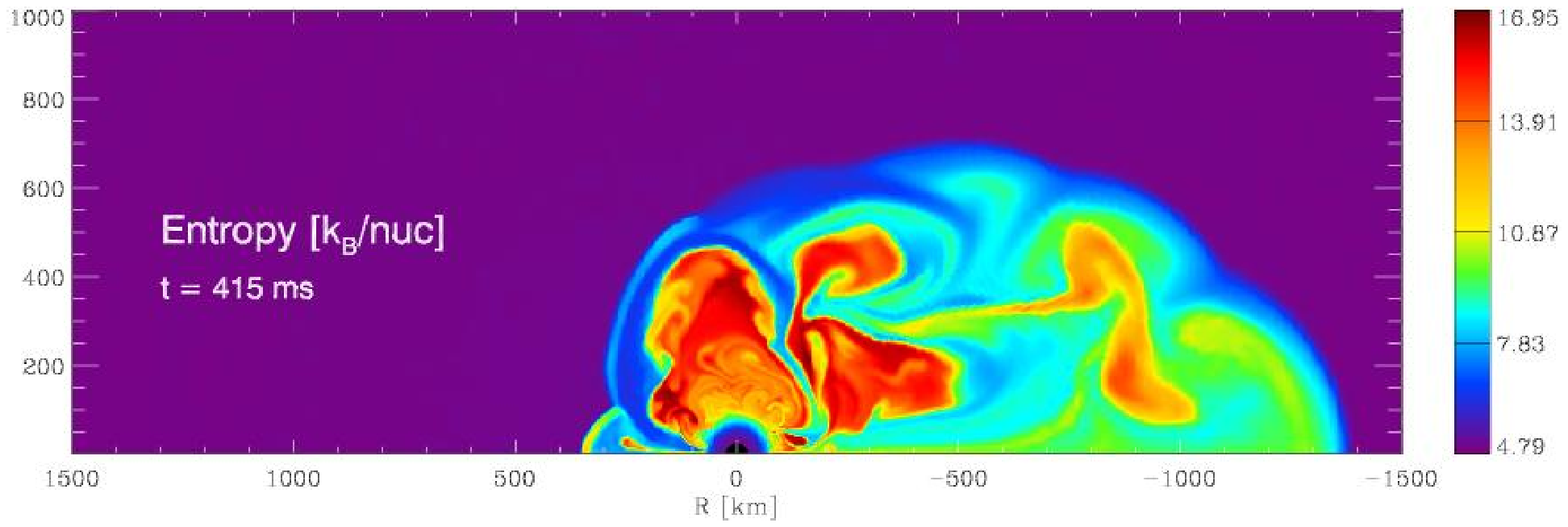}
\end{center}
\vspace{-5mm}
\begin{center}
\includegraphics[width=1.0\textwidth]{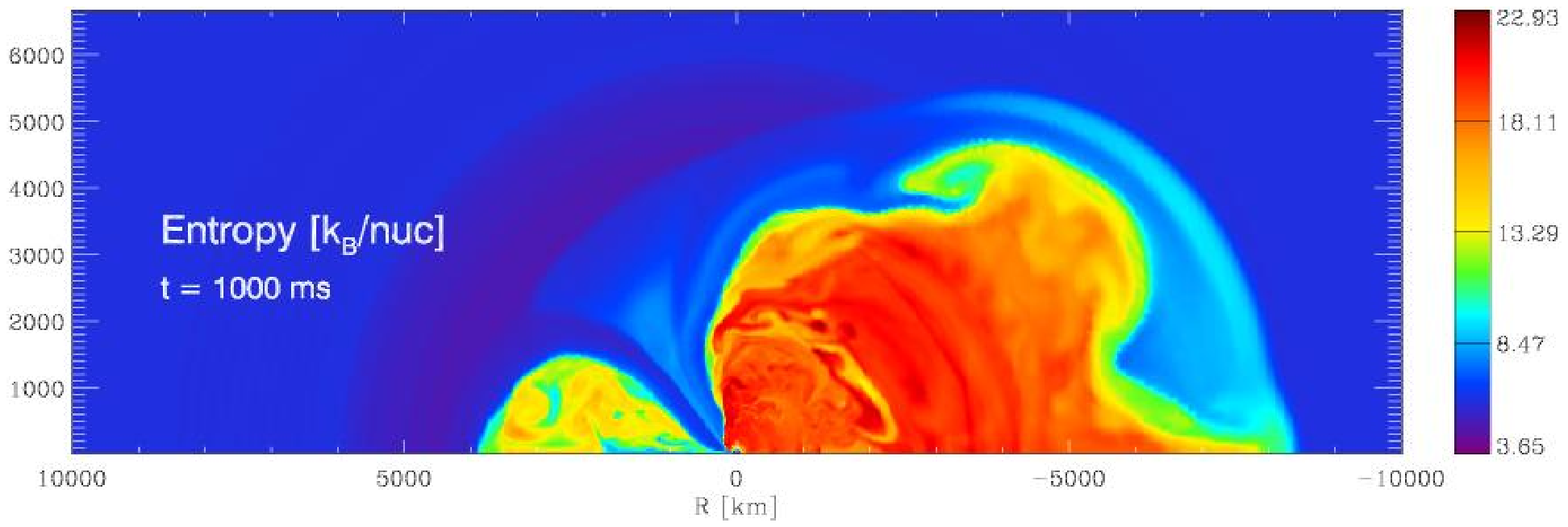}
\end{center}
\caption[]{Explosion that is driven by neutrino-energy deposition 
in combination with convective overturn in the region
behind the supernova shock. The anisotropy of the neutrino-
and shock-heated ejecta is growing in time and becomes very
large due to an increasing contribution of the $m=0$, $l=1$ 
mode in the convective pattern. The snapshots (from top to 
bottom) show the entropy distribution (values between
about 4 and 23$\,k_{\mathrm B}$ per nucleon) at post-bounce times
$t_{\mathrm{pb}} = 245\,$ms, 415$\,$ms, 
and 1000$\,$ms. 
Note that the radial scales of the figures differ.
The neutron star is at the origin of
the axially symmetric (2D) grid and plays the role of an
isotropic neutrino ``light bulb''~\cite{ple02}.} 
\label{fig:monomode}
\end{figure}

Rotation plus magnetic fields were proposed as the 
``most obvious'' way to break the spherical symmetry and to
explain the global asphericity of core-collapse 
supernovae~\cite{whemei02,akiwhe02,whe02}.
It was argued that current numerical calculations may be 
missing a major ingredient necessary to yield explosions.
A proper treatment 
of rotation {\em and} magnetic fields may be necessary to 
fully understand when and how collapse leads to explosions.
Of course, this might be true. But a confirmation or 
rejection will require computer models with ultimately the full
physics. 

It must be stressed, however, that current observations
do not necessitate such conclusions and hydrodynamical
simulations suggest other possible explanations. 
Strong convection in the
neutrino-heating region behind the supernova shock can account
for huge anisotopies of the inner supernova ejecta, even 
without invoking rotation. If the explosion occurs quickly,
much power remains on smaller scales until the expansion sets in
and the convective pattern gets frozen in. If, in contrast, the 
shock radius grows only very slowly and the explosion is delayed
for several 100$\,$ms after bounce, the convective flow can
merge to increasingly larger structures. 
In two-dimensional (2D) hydrodynamic calculations including 
cooling and heating by neutrinos between the neutron star and
the shock (with parameter choices for a central, {\em isotropic} 
neutrino ``light bulb'' which enabled explosions),
Plewa et al.~\cite{ple02} found situations
where the convective pattern revealed a contribution of the $l=1$,
$m=0$ mode that was growing with time and was even dominant at
about one second after bounce (Fig.~\ref{fig:monomode}). 
Herant~\cite{her95} already speculated about such a possibility.
Certainly three-dimensional (3D) calculations of the full sphere
(and without the coordinate singularity on the axis of the
spherical grid) are indispensable to convincingly demonstrate 
the existence of this phenomenon\footnote{Another interesting 
possibility
was pointed out by A.~Mezzacappa at this conference. He showed
results of 2D and 3D calculations performed in collaboration 
with J.M.~Blondin~\cite{blomez02}, which revealed hydrodynamical 
instabilities growing to large-scale modes in the flow behind 
the accretion shock even in the absence of neutrino heating.}.

\section{Do Neutrino-Driven Explosions Work?}
\label{neutrinoexplosions}

Spherically symmetric simulations with the current input
physics (neutrino interactions and the equation of state of
dense matter) do not yield explosions by the neutrino-heating
mechanism. There is no controversy about that. All computations
are in agreement, independent of Newtonian or relativistic
gravity and independent of the neutrino transport being treated
in an approximate way by flux-limited diffusion 
methods (e.g.,~\cite{myrblu87,myrblu89,bru93,brunis01}) or
very accurately by solving the frequency- and angle-dependent
Boltzmann transport 
equation~\cite{ramjan00,mezlie01,liemez01,liemes02}.

Whether neutrinos succeed in reviving the stalled shock
depends on the efficiency 
of the energy transfer to the postshock layer, which in turn
increases with the neutrino luminosity and the hardness of
the neutrino spectrum. 
Wilson and collaborators~\cite{wilmay88,wilmay93,maytav93,totsat98}
have obtained explosions in one-dimensional (1D) simulations for more 
than ten years now. In these models it is, however, {\em assumed} that
neutron-finger convection in the hot neutron star boosts the 
neutrino luminosities. Moreover, Mayle et~al.~\cite{maytav93}
used a special equation of state with a high abundance of 
pions in the nuclear matter, which again leads to higher neutrino
fluxes from the neutron star and thus to enhanced 
energy-deposition behind the shock. Both assumptions
are not generally accepted.

Two-dimensional~\cite{herben92,herben94,shiyam94,burhay95,janmue96,mezcal98:ndconv,shiebi01}
and 3D simulations \cite{shiyam93,frywar02} have shown
that the neutrino-heating layer is unstable to
convective overturn. The associated effects have a very helpful 
influence and can lead to explosions even in cases where 
spherical models fail. 
In the multi-dimensional situation downflows of cooler,
low-entropy matter that has fallen through the shock, 
coexist with rising bubbles of high-entropy, neutrino-heated 
gas. On the one hand, the downflows take cool material close
to the gain radius where it absorbs energy readily from the
intense neutrino fluxes. On the other hand, the rising bubbles
allow heated matter to expand and cool quickly, thus reducing
the energy loss by the reemission of neutrinos. They also
increase the postshock pressure and hence push
the shock farther out. This in turn enlarges the gain layer
and thus the gas mass which can accumulate in the 
neutrino-heating region.
It also means that the gas stays longer in the gain layer, in
contrast to one-dimensional models where the matter behind the 
accretion shock has negative velocity and is quickly advected
down to the cooling layer. When the gas arrives there, 
neutrino emission
sets in and extracts again the energy which had been absorbed 
from neutrino heating shortly before. Due to the combination of
all these effects postshock convection enhances the efficieny
of the neutrino-heating mechanism. Therefore the 
multi-dimensional situation is {\em generically different} from 
the spherically symmetric case.

Nevertheless, the existence of convective overturn in the
neutrino-heating layer does not guarantee
explosions~\cite{janmue96,mezcal98:ndconv}.
For insufficient neutrino heating the threshold to an 
explosion will not be overcome. Since neutrinos play a crucial
role, an accurate description of the neutrino physics ---
transport and neutrino-matter interactions --- is indispensable
to obtain conclusive results about the viability of the 
neutrino-driven mechanism. All published multi-dimensional 
explosion models, however, have employed crude approximations
or simplifications in the treatment of neutrinos.

\begin{figure}[htb!]
\begin{center}
\includegraphics[width=0.93\textwidth,clip=]{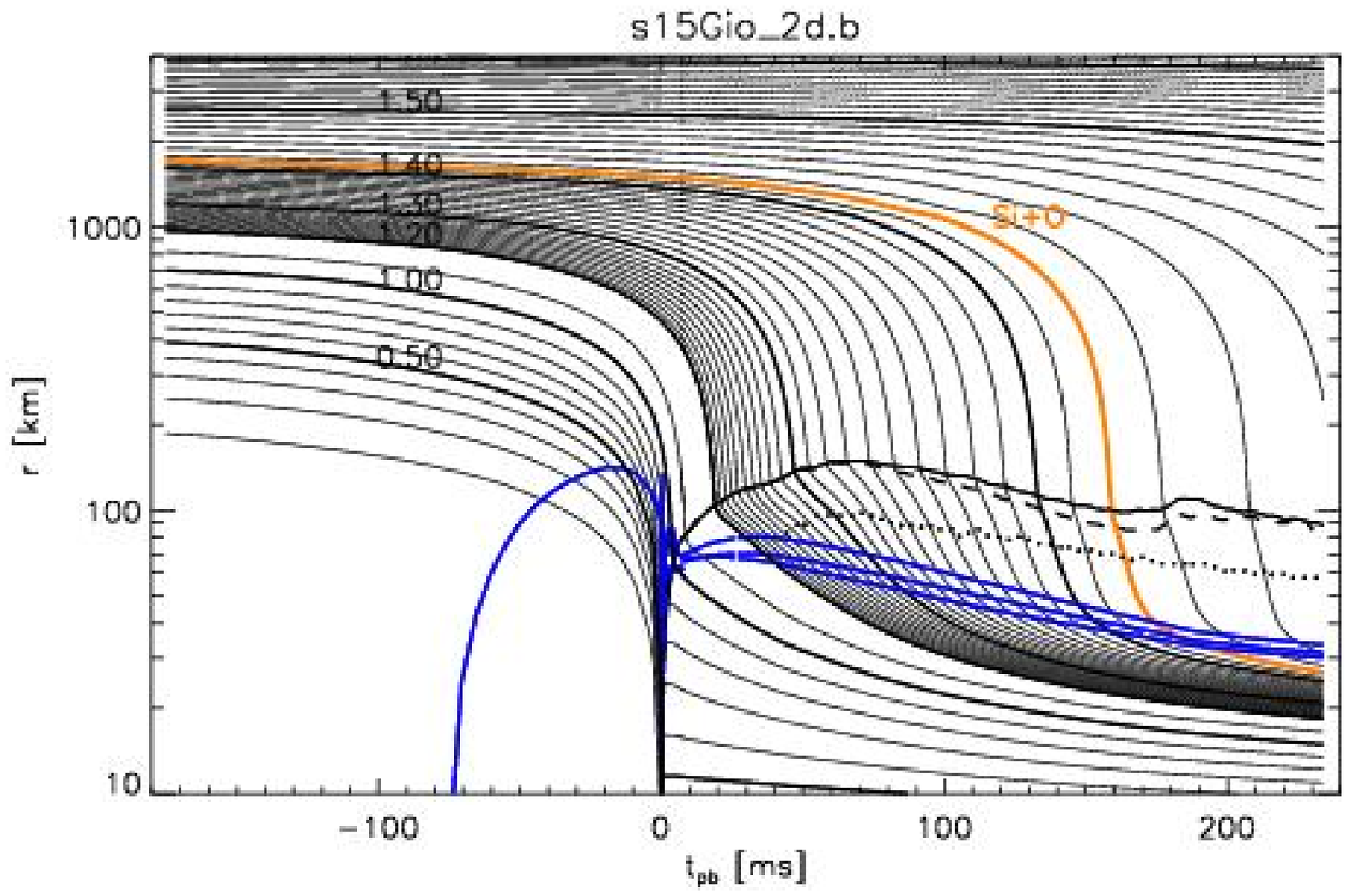}
\end{center}
\vspace{-3mm}
\begin{center}
\includegraphics[width=0.93\textwidth,clip=]{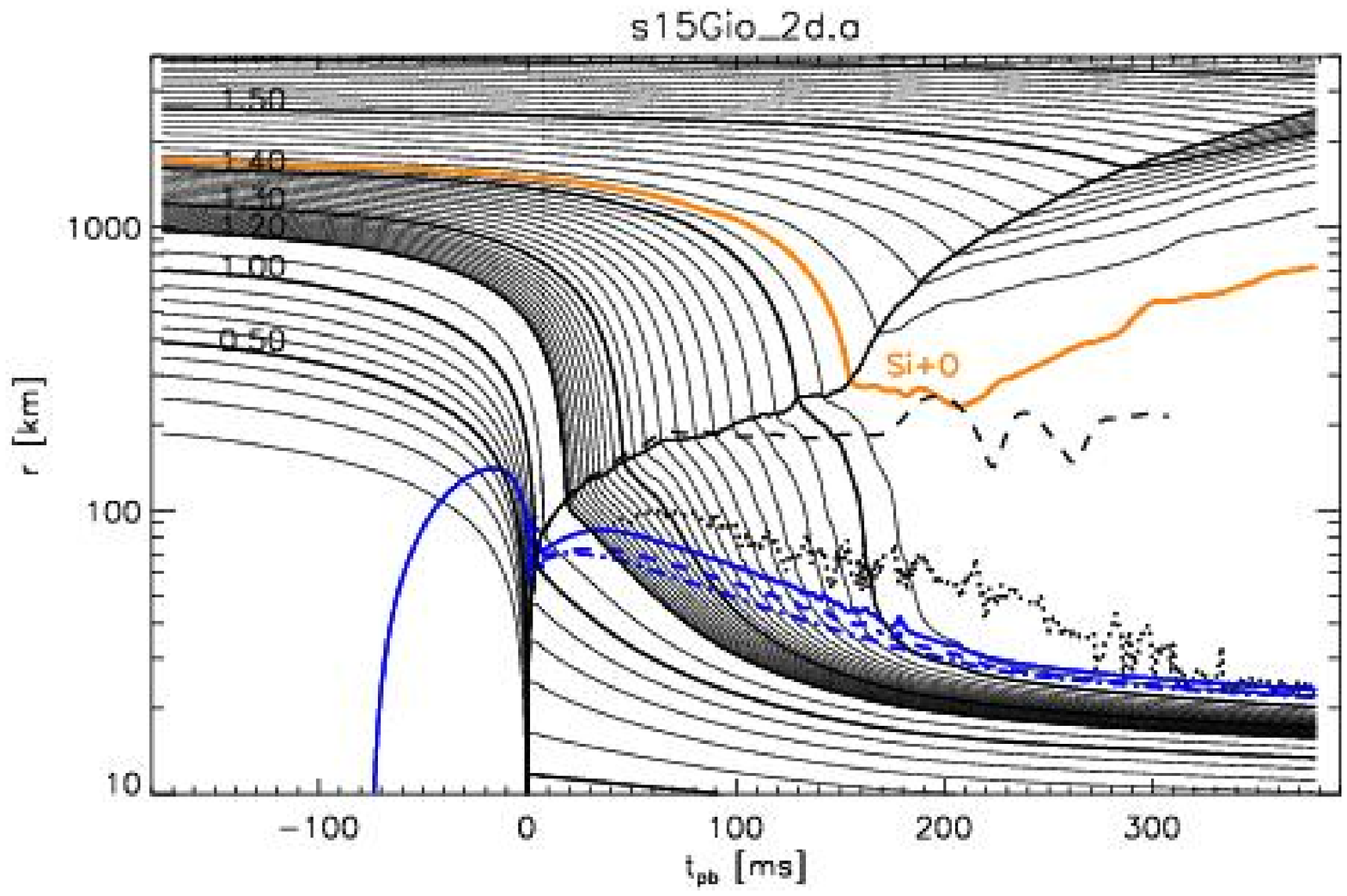}
\end{center}
\caption[]{Trajectories of mass shells (time being
normalized to bounce) for the
non-exploding (top) and the exploding 2D model.
In the latter case one can see the
shock starting a rapid expansion at about 150$\,$ms after
bounce. The dashed lines indicate the shock positions in
the corresponding 1D simulations, where no explosions were
obtained. The angle-averaged gain radius is given by the
dotted line, and the neutrinospheres of $\nu_e$, $\bar\nu_e$ 
and the heavy-lepton neutrinos are also marked.} 
\label{fig:massshells}
\end{figure}

\section{A New Generation of 2D Supernova Simulations}
\label{sec:results}

In order to take a next step of improvement in supernova
modelling, we have coupled a 
new Boltzmann code for the neutrino transport to the 
PROMETHEUS hydrodynamics program, which allows for 
spherically symmetric as well as multi-dimensional
simulations~\cite{ramjan02}.
Below we present some results of our first 2D supernova
simulations with this new code, which we named
MuDBaTH ({\bf Mu}lti-{\bf D}imensional {\bf B}oltzm{\bf a}nn
{\bf T}ransport and {\bf H}ydrodynamics).

\subsection{Technical Aspects and Input Physics}
\label{sec:code}
 
The Boltzmann solver scheme
is described in much detail in Ref.~\cite{ramjan02}. The
integro-differential character of the Boltzmann equation is
tamed by applying a variable Eddington factor closure to the
neutrino energy and momentum equations (and the simultaneously
integrated first and second order moment equations for neutrino
number). For this purpose the variable Eddington factor
is determined from the solution of the Boltzmann equation, and
the system of Boltzmann equation and its moment equations is
iterated until convergence is achieved. Employing this scheme in
multi-dimensional simulations in spherical coordinates, we solve the
(one-dimensional)
moment equations on the different angular bins of the numerical grid but
calculate the variable Eddington factor only once on an angularly
averaged stellar background. 
We point out here that it turned out to be
necessary to go an important step beyond this simple ``ray-by-ray''
approach. Physical constraints, namely the conservation of lepton
number and entropy within adiabatically moving fluid elements,
and numerical requirements, i.e., the stability of regions
which should not develop convection according to a mechanical stability
analysis, make it necessary to take into account the coupling of
neighbouring rays at least by lateral advection terms and neutrino
pressure gradients~\cite{burram02}.

General relativistic effects are treated only approximately in
our code~\cite{ramjan02}. The current version contains a modification
of the gravitational potential by including correction terms due to
pressure and energy of the stellar medium and neutrinos, which
are deduced from a comparison of the Newtonian and relativistic
equations of motion. The neutrino transport contains gravitational   
redshift and time dilation, but ignores the distinction between
coordinate radius and proper radius. This simplification is
necessary for coupling
the transport code to our basically Newtonian hydrodynamics.

As for the neutrino-matter interactions, we discriminate between
two different sets of input physics. On the one hand we have
calculated models with conventional (``standard'') neutrino
opacities, i.e.,
a description of the neutrino interactions which follows closely
the one used by Bruenn and Mezzacappa and collaborators
\cite{bru85,mezbru93:coll,mezbru93:nes}. It assumes
nucleons to be uncorrelated, infinitely massive scattering
targets for neutrinos. In these reference runs we have
usually also added
neutrino pair creation and annihilation by nucleon-nucleon
bremsstrahlung \cite{hanraf98}. Details of our
implementation of these neutrino processes can be found 
in~\cite{ramjan02}.

A second set of models was computed with
an improved description of neut\-rino-\-matter interactions.
Besides including nucleon thermal motions and recoil, which
means a detailed treatment of the reaction kinematics
and allows for an accurate evaluation of nucleon phase-space
blocking effects, we take into account nucleon-nucleon
correlations (following Refs.~\cite{bursaw98,bursaw99}),
the reduction of the nucleon effective mass, and the
possible quenching of the axial-vector coupling
in nuclear matter \cite{carpra02}. In addition, we have
implemented weak-magnetism corrections as described in
Ref.~\cite{hor02}. The sample of neutrino processes was
enlarged by also including scatterings of muon and tau
neutrinos and antineutrinos off electron neutrinos and
antineutrinos and pair annihilation reactions between
neutrinos of different flavors
(i.e., $\nu_{\mu,\tau} + \bar\nu_{\mu,\tau}
\longleftrightarrow \nu_e + \bar\nu_e$; \cite{burjan02:nunu}).

Our current supernova models are calculated with the nuclear
equation of state of Lattimer and Swesty \cite{latswe91},
which we suitably extended to lower densities~\cite{ramjan02}.

\begin{figure}[htb!]
\begin{center}
\includegraphics[width=1.0\textwidth]{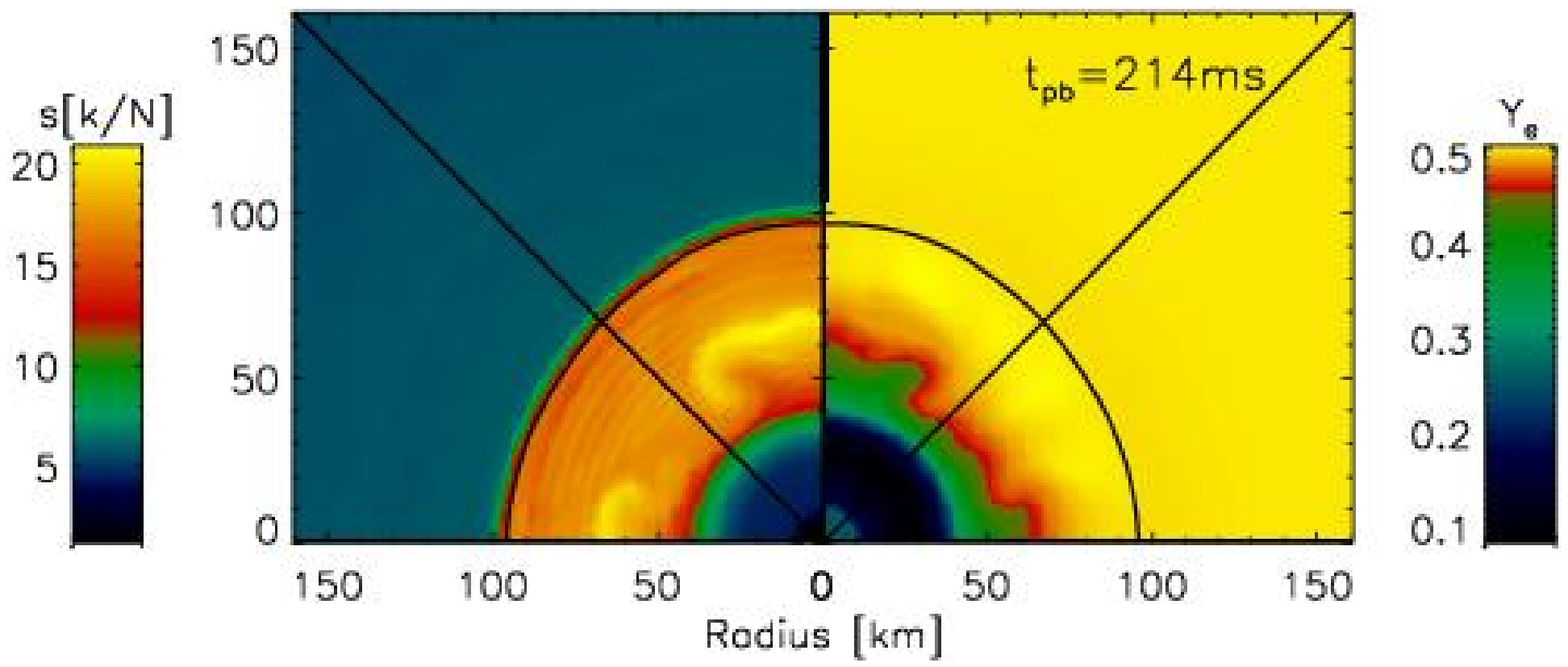}
\end{center}
\vspace{-5mm}
\begin{center}
\includegraphics[width=1.0\textwidth]{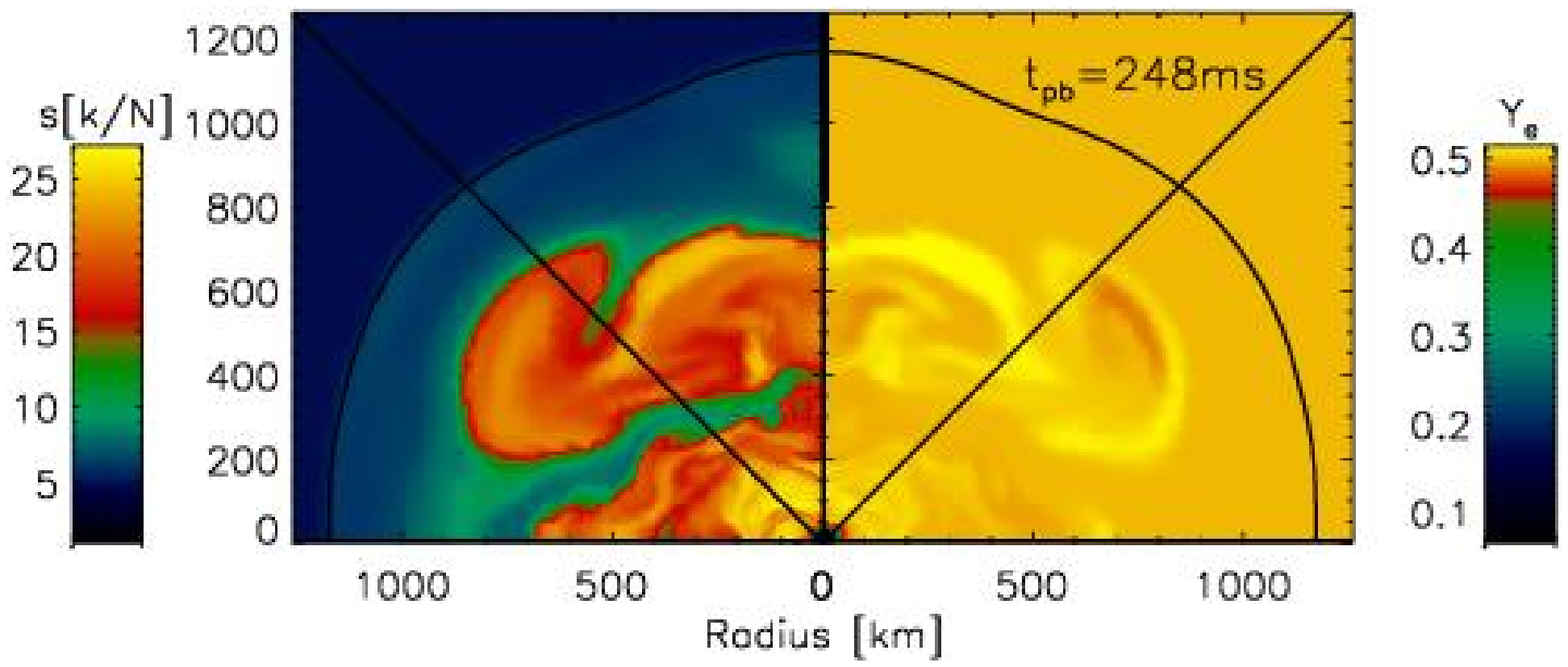}
\end{center}
\caption[]{Convection in the neutrino-heating region for
the non-exploding 2D model (Model~s15Gio\_2d.b, top) 
and the exploding one (Model~s15Gio\_2d.a) at the
post-bounce times indicated in the plots. The figures show
the entropy distribution (left) and the
electron fraction (proton-to-baryon ratio). A wedge of
$\pm 43.2^{\mathrm o}$ around the equatorial plane (marked
by the diagonal solid lines) of the spherical coordinate
grid was used for the computations.}
\label{fig:snapshots}
\end{figure}

\subsection{Models and Results}
\label{sec:models}

None of our spherically symmetric simulations, neither 
with the standard nor with the improved description of 
neutrino opacities, has produced an explosion. 
A compilation of a subset of our calculations which we
did for a 15$\,$M$_{\odot}$ progenitor star, Model s15s7b2,
provided to us by S.~Woosley, can be found in 
Refs.~\cite{janbur02a,burram02}. 
Here we discuss only two 2D runs
(Models s15Gio\_2d.a and s15Gio\_2d.b), which both were
performed with our approximation of relativistic effects 
and the state-of-the-art improvement of neutrino-matter 
interactions (cf.~Sect.~\ref{sec:code}). We used a spherical
coordinate grid with 32 equidistant zones within an angular
wedge from $-43.2^{\mathrm{o}}$ to $+43.2^{\mathrm{o}}$
around the equatorial plane and assumed periodic conditions
at the boundaries.

Both 2D simulations differ only in
one important aspect: In Model~s15Gio\_2d.a the velocity 
dependent (Doppler shift and aberration)
terms in the neutrino momentum equation (and the corresponding
terms in the Boltzmann equation for the antisymmetric average
of the specific intensity; see Ref.~\cite{ramjan02}) were
omitted. These terms are formally of order $v/c$ and are
small for low velocities. 

This simplification of the
neutrino transport, however, has a remarkable consequence:
The model with the most complete implementation of the 
transport equations, Model~s15Gio\_2d.b, fails to explode.
In case of Model~s15Gio\_2d.a, however, 
the stalled shock is successfully revived by
neutrino heating because very strong convection can
develop in the gain region\footnote{At the
time of the conference, we had just this exploding 2D run 
and made a preliminary announcement of the success of this
model. A later 2D computation with the full neutrino moment
equations (Model~s15Gio\_2d.b) then turned out to produce a 
dud.}. The time evolution of both models is displayed 
by the trajectories of mass shells in
Fig.~\ref{fig:massshells}. 

The reason for this dramatic difference is the following.
Some of the velocity dependent terms (those in which
derivatives with respect to the neutrino energy do not
show up) in the neutrino momentum equation
have a simple formal interpretation: In regions
with mass infall (negative velocity) they effectively
act like a reduction
of the neutrino-medium interaction on the right hand
side of this equation. The changes can be 10\% or more
for neutrino energies in the peak of the spectrum,
depending also on time, radius, and the size of the
postshock velocities. As a consequence, the neutrino flux
streams more readily and the comoving-frame neutrino (energy)
density is decreased.
This is associated with somewhat larger neutrino losses 
in the cooling layer around the neutrinosphere and a 
significantly reduced neutrino heating between 
gain radius and shock. 

Although the differences are moderate
(10--30\%, depending on the quantity)
the accumulating effects during the first 80$\,$ms after bounce
clearly damp the shock expansion and finally lead to a dramatic 
shock recession after the initial phase of expansion.
Before this happens postshock convection has not become strong
enough to change the evolution. With the onset of contraction, 
the postshock velocities decrease (become more negative)
quickly, neutrino-heated
matter is rapidly advected inward below the gain radius and 
loses its energy by reemission of neutrinos. The gain region
shrinks to a very narrow layer, a fact which suppresses 
the convective activity lateron. This is 
demonstrated by Fig.~\ref{fig:snapshots} where convection
is weak in Model~s15Gio\_2d.b but very strong in 
Model~s15Gio\_2d.a.

Due to a combination of unfavorable effects and a continuously
amplifying negative trend, Model~s15Gio\_2d.b remains below the 
explosion threshold while Model~s15Gio\_2d.a is just above 
that critical limit. In the vicinity of the threshold the
long-time evolution of the collapsing stellar core depends
very sensitively on ``smaller details'' of the neutrino 
transport.

\section{Conclusions and Outlook}
\label{sec:conclusions}

Our 2D models with a Boltzmann solver for the 
neutrino transport have considerably reduced the
uncertainties associated with the treatment of the
neutrino physics in previous multi-dimensional 
simulations. With the most complete implementation of
the transport physics we could not obtain explosions.
This result suggests that the neutrino-driven
mechanism fails with the employed input physics,
at least in case of the considered 15$\,$M$_{\odot}$
star. We do not think that the remaining
uncertainties in our simulations (mainly the 
approximate treatment of general relativistic effects)
are likely to jeopardize this conclusion. A comparison
with fully relativistic one-dimensional calculations
(Liebend\"orfer, personal communication~\cite{lieetal02}) 
is very encouraging. Because of the remarkable similarity 
of the shock trajectories of different progenitors 
in spherical symmetry~\cite{liemes02}, it is likely
that our negative conclusion is also valid for other
pre-collapse configurations with a similar structure.
Significant star-to-star variations of the 
progenitor properties with a non-monotonic dependence 
on the stellar mass~\cite{wooheg02},
however, suggest that 
multi-dimensional core-collapse simulations of a
larger sample of progenitors are needed before one can 
make final, more generally valid statements.
The supernova problem is highly nonlinear 
and surprises may lurk behind every corner.

It would therefore be premature to conclude that the
neutrino-driven mechanism fails and that not even 
postshock convection can alter this unquestioned 
outcome of all current spherical models. Besides 
studying other progenitors with multi-dimensional 
simulations, one should also investigate the effects
of rotation and the influence of different high-density
equations of state on the long-time post-bounce evolution 
and the neutrino-heating phase in a supernova. There is
considerable uncertainty associated with the poorly
known physics in the nuclear and supranuclear medium.

Our successfully exploding 2D model, Model~s15Gio\_2d.a, 
at least demonstrates that simulations which include the
effects of postshock convection are rather close
to an explosion. Therefore modest changes of the neutrino 
emission and transport seem to be already sufficient to push 
them beyond the critical
threshold. The properties of the explosion in this case are
very encouraging and may support one's belief in the basic 
viability 
of the delayed explosion mechanism. At 380$\,$ms after bounce 
the shock has arrived at a radius of more than 2500$\,$km and
is expanding with about 10000$\,$km/s.
The explosion of this model does not seem to become very
energetic. It is only $\sim 4\times 10^{50}\,$erg at that 
time, but still increasing. This may not be a serious problem
if one recalls the large spread of energies
of observed supernovae (Supernova~1999br, for example, 
is estimated to have an ejecta mass of $14\,$M$_{\odot}$
and an explosion energy of about $6\times 10^{50}\,$erg~\cite{ham02}).

Since the explosion starts rather late (at $\sim 150\,$ms post 
bounce), the proto-neutron star has accreted enough matter to have  
attained an initial baryonic mass of 1.4$\,$M$_{\odot}$. Therefore
our simulation does not exhibit the problem of previous
successful multi-dimensional calculations which produced 
neutron stars with masses on the lower side of plausible
values ($\sim 1.1\,$M$_{\odot}$).
Also another problem of published explosion models
(e.g.,~\cite{herben94,burhay95,janmue96,fry99})
has disappeared: The ejecta mass with $Y_e \la 0.47$
is less than $10^{-4}\,$M$_{\odot}$,
thus fulfilling a constraint pointed out by
Hoffman et.~al.~\cite{hofwoo96} for supernovae
if they should not overproduce the $N=50$ (closed neutron
shell) nuclei, in particular $^{88}$Sr, $^{89}$Y and $^{90}$Zr,
relative to the Galactic abundances. Of course, final statements
about explosion energy, ejecta composition, and the neutron star
mass (which may grow by later fallback, especially when the explosion
energy remains low) require to follow the explosion for a longer time.

\bigskip\noindent
{\small
{\bf Acknowledgements:} We are grateful to K.~Takahashi for
providing routines to calculate the improved neutrino-nucleon
interactions, and to C.~Horowitz for correction formulae for the
weak magnetism.
We also thank M.~Liebend\"orfer for making output data of his 
simulations available to us for comparisons.
The Institute for Nuclear Theory at the University
of Washington is acknowledged for its hospitality and the
Department of Energy for support during a visit of the Summer Program
on Neutron Stars, during which most of the work leading to 
Fig.~\ref{fig:monomode} was done. HTJ, RB and MR
are grateful for support by the Sonderforschungsbereich
375 on ``Astroparticle Physics'' of the Deutsche Forschungsgemeinschaft.
TP was supported in part by the US Department of Energy
under Grant No. B341495 to the Center of Astrophysical Thermonuclear
Flashes at the University of Chicago, and in part by the grant
2.P03D.014.19 from the Polish Committee for Scientific Research.
He performed his simulations on the CRAY SV1-1A at the 
Interdisciplinary Centre for Computational Modelling in Warsaw.
The 2D simulations with Boltzmann neutrino transport were only 
possible because a node of the   
new IBM ``Regatta'' supercomputer was dedicated to this project
by the Rechenzentrum Garching. Computations were also
done on the NEC SX-5/3C
of the Rechenzentrum Garching, and on the CRAY T90 and CRAY
SV1ex of the John von Neumann Institute for Computing (NIC) in J\"ulich.
}

\end{document}